\documentclass[aps,prl,twocolumn,superscriptaddress,showpacs]{revtex4-1}

\usepackage{graphicx}
\usepackage{dcolumn}
\usepackage{amsmath,bm}
\usepackage{color}%

\begin{document}

\title{Carbon nanotubes with atomic impurities on boron nitride sheets under applied electric fields}

\author{Seoung-Hun Kang}
\affiliation{Department of Physics and
             Research Institute for Basic Sciences,
             Kyung Hee University, Seoul, 130-701, Korea}

\author{Gunn Kim}
\email[e-mail: ]{gunnkim@sejong.ac.kr}
\affiliation{Department of Physics, 
             Graphene Research Institute and 
             Institute of Fundamental Physics,
             Sejong University, Seoul, 143-747, Korea}

\author{Young-Kyun Kwon}
\email[e-mail: ]{ykkwon@khu.ac.kr}%
\affiliation{Department of Physics and
             Research Institute for Basic Sciences,
             Kyung Hee University, Seoul, 130-701, Korea}

\date{\today}

\begin{abstract}
We perform first-principles calculations to investigate 
the structural and electronic properties of metal-doped (10, 0) carbon nanotubes (CNTs) 
on a single hexagonal boron nitride (hBN) sheet
in the presence of an external electric field.
We consider K, Cl and Ni atoms as dopants to study the dependence of the electronic properties of the CNT 
on doping polarity and concentration. 
The electric field strength is varied from $-$0.2 V/\AA~to $+$0.2 V/\AA~to explore 
the effects of an external electric field on the electronic structures. 
Although the electronic energy bands of the hBN 
sheet are modified in accordance with the field strength, its 
electronic state in the valence or conduction band does not touch the Fermi level under the field strength 
considered. We conclude that the hBN as a substrate does not modify the 
electronic structure of the CNT, thereby leading to improvements in the device performance, 
compared with that of devices based on conventional substrate materials such as SiO$_2$.
\end{abstract}





\maketitle

%


The search for substrates to improve the properties of the nanoelectronic devices has been a 
crucial focus of research since the time of fabrication and use of such devices.
Thus far, metals~\cite{{GRARU},{GRAIR}}, mica~\cite{GRAMICA}, 
SiC\cite{{SIC1},{SIC2}} and SiO$_2$~\cite{{SIO21},{SIO22},{SIO23},{SIO24},{SIO25}} 
have been used as substrates. For graphene-based devices, SiO$_2$ is 
most commonly used as a substrate. Although the use of SiO$_2$ provides many advantages,
its primary drawback is that charge density fluctuations 
induced by the presence of impurities reduce electronic mobility in the device~\cite{{REDM},{REDM1}}.

Recently, the use of hexagonal boron nitride (hBN) as a substrate [Figure 1(a)] for graphene 
devices has attracted considerable interest because compared with conventional substrates, 
graphene on the hBN substrate exhibits increased mobility~\cite{{IMMO},{HQGRA}} and further, 
significant improvements has been observed in quantum Hall measurements~\cite{HQGRA}. 
According to the studies based on scanning tunneling
microscopy measurements~\cite{{STM1},{STM2}}, compared with SiO$_2$ substrates, 
graphene on hBN provides an extremely flat surface that significantly reduces electron-hole puddles.
In addition to the advanctages of providing an atomically smooth surface (relatively free of dangling bonds
and charge traps), a large band gap (${\approx}$6 eV~\cite{BG}), chemical inertness,
and a low density of charged impurities, hBN sheets 
also interestingly exhibit an unusual electronic structure 
not observed in most wide bandgap materials. 

\begin{figure}[b]
\includegraphics[width=0.9\columnwidth]{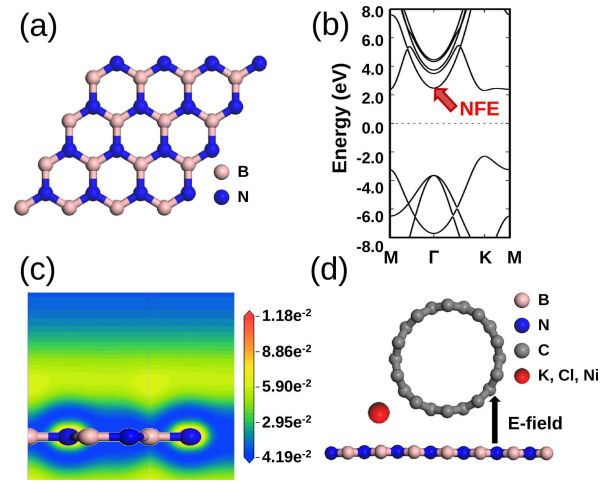}
\caption{(Color online) The NFE state of the BN sheet and the doped CNT
on the BN sheet under an applied electric field. (a) A free-standing
4 $\times$ 4 hBN sheet. (b) The NFE state at ${\Gamma}$ in the band structure of
the BN sheet. (c) The contour plot of the NFE state of the BN sheet shows
electron accumulation (red) and depletion (blue).
(d) Side view of a metal-doped (10, 0) CNT on the BN sheet under an applied electric field
along the $+z$ direction. White, blue, gray, and red balls represent boron,
nitrogen, carbon, and metal dopant atoms, respectively.
In (d), the arrow indicates the direction of the external electric field.
\label{Fig1}}
\end{figure}

Previous studies have reported that the nearly free electron (NFE) state~\cite{NFE} exists in the conduction band minimum (CBM) 
of hBN at the ${\Gamma}$ point ($\vec{k}=0$), as shown in Figure 1(b). Figure 1(c) shows 
the NFE state at ${\Gamma}$. It has an electronic density about 3 \AA~above the hBN layer (the yellow region). 
If a carbon nanotube (CNT) is placed on the BN sheet, the charge can move
from the CNT to hBN. Owing to the presence of an external electric field, doping and so on, 
the free-electron-like energy dispersion of the hBN substrate may lead to
unwanted electrical conduction (leakage current) through the hBN substrate.

In this Letter, we report the structural and electronic properties of  
doped (10, 0) CNTs on an hBN sheet in the presence of an 
external electric field; the structural and electronic properties are obtained 
by using first-principles pseudopotential density-functional methods. 
We consider three different types of dopants (K, Cl, and Ni) 
to study the electronic property dependence on the doping polarity, 
and we calculate the projected density of states (PDOS) to determine whether the electronic states originating from 
the hBN substrate equal the Fermi level ($E_F$), causing some problems in CNT devices.
Finally, we show that although the electronic energy bands of the hBN sheet
are modified in accordance with the field strength, its electronic state in the valence or conduction band does not equal 
$E_F$ under the field strength considered. Our result suggests that 
hBN as a substrate does not modify the electronic structure of the CNT, thereby leading to  
improvements in device performance, when comparing the performance of CNT-based devices 
on the hBN substrate with conventional substrate materials such as SiO$_2$. 
However, residual scattering and electrical conduction could occur in the case of doped-CNTs placed on the hBN substrate
when the top gate voltage is applied.




We carried out first-principles calculations using the Vienna Ab-initio Simulation 
Package (VASP)~\cite{VASP}. The projector-augmented wave pseudopotentials
were employed~\cite{PAW}. The exchange-correlation functional was treated within 
the spin-polarized local density approximation (LDA) in the form of 
Ceperley-Alder-type parameterization~\cite{LDACA}. The cutoff energy 
for the planewave basis expansion was chosen to be 450 eV, and the atomic 
relaxation was continued until the Helmann-Feynman forces acting on atoms were 
smaller than 0.03 eV/\AA. For more precise calculations, we included the dipole correction.

The BN sheet was represented by a slab of a single-layered BN sheet and a vacuum region
of 14~\AA. We arranged the CNT and the BN sheet on a tetragonal lattice, with one side of the 
unit cell being equal to the lattice constant of the zigzag CNT. The other sides of 
the supercell were set to a length of 17.5~\AA, about twice the value of the CNT 
diameter, to avoid intertube interaction. Here, the B-N bond length was calculated
to be 1.44~\AA~(the corresponding lattice constant of the BN sheet was 2.50~\AA), 
which is reasonably close to the experimental value of 2.51~\AA. The Brillouin zone
was sampled using a $\Gamma$-centered 10 $\times$ 1 $\times$ 1 $k$-point mesh for
all unit cells. The electronic levels were convoluted using Gaussian broadening
with a width of 0.05 eV to obtain the DOS.

To investigate the influence of adsorbed charge dopants, we chose potassium (K) and chlorine (Cl) atoms. ͑
Potassium is a member of the alkali metal family and it tends to donate one electron to the CNT or the graphene sheet~\cite{{KCNT1},{KCNT2}},
whereas chlorine is a member of the halogen family and it tends to accept one electron from the CNT or the graphene sheet~\cite{{ClCNT1},{ClCNT2}}.
On the other hand, nanoparticulate nickel is often used as a catalyst to prepare CNTs. If the Ni nanoparticles are not completely
removed from the CNT surfaces after the growth, the CNT may contain nickel impurity
atoms on its surface~\cite{NiCNT}. Therefore, we considered a nickel-doped CNT for this investigation.


\begin{figure}[t]
\includegraphics[width=0.9\columnwidth]{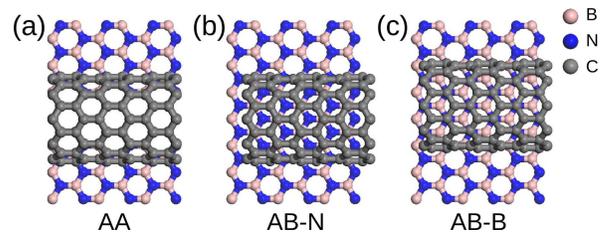}
\caption{(Color online) 
Three possible sites of our supercell model with the CNT adsorbed on the BN sheet. 
(a) Top view for AA-stacking. (b) Top view for AB-N stacking. 
(c) Top view for AB-B stacking. 
White, blue and gray balls represent boron, nitrogen and carbon atoms, respectively. 
\label{Fig2}}
\end{figure}

To examine whether the NFE state originating from the hBN 
substrate touch $E_F$, we applied an external uniform electric field to the doped CNT. Firstly,  
we studied the effect of an external electric field on the electronic properties
of an undoped (10, 0) CNT on the BN sheet substrate. 
The unit cell dimension of the BN sheet was chosen to be equal to one unit of the CNT.
Figure 2 shows three possible sites of our 
supercell models with the CNT adsorbed onto the BN sheet, namely, the top site of each 
atom (AA-stacking), the hollow site formed by the B3N3 hexagon and the top site 
of a B atom (AB-B stacking), and the hollow site formed by the B3N3 hexagon and the top 
site of a N atom (AB-N stacking). The energetically favorable site was found to be AB-N stacking.

\begin{figure*}[t]
\includegraphics[width=1.7\columnwidth]{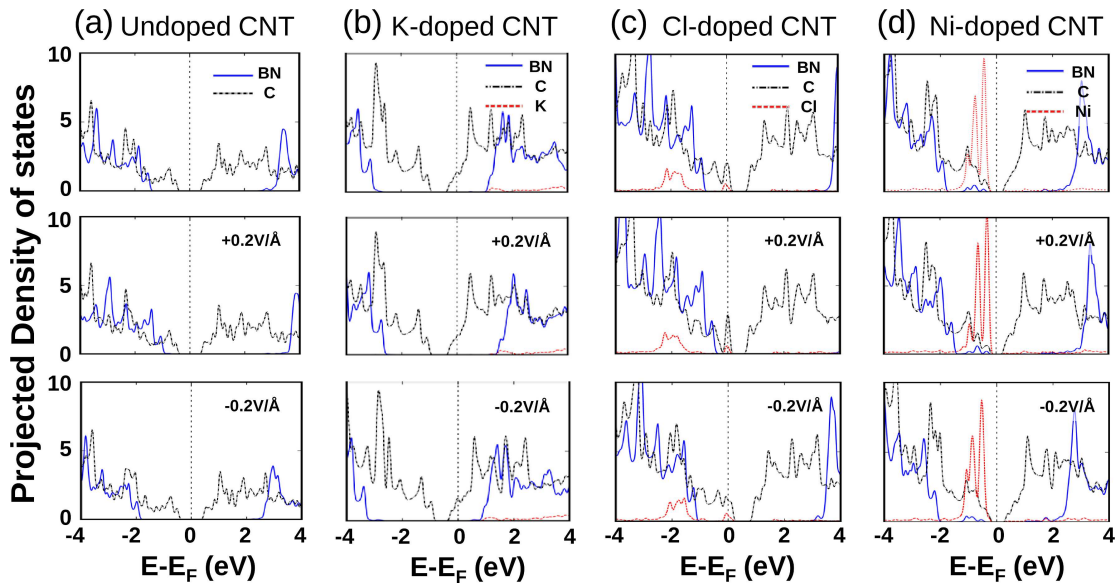}
\caption{(Color online) 
Undoped and doped (10, 0) CNTs placed on the BN sheet in the presence of an external electric 
field. (a) The PDOSs of the CNT and the BN sheet for the undoped CNT on the BN sheet
with and without the field. The field direction is along the $z$ direction, which is normal to the BN sheet.
(b) The PDOS of a potassium-doped CNT on the BN sheet under an applied electric field. 
(c) The PDOS of a chlorine-doped CNT on the BN sheet 
under an applied electric field.
(d) The PDOS of the nickel-doped CNT on the BN sheet under an applied electric
field. The Gaussian broadening for the PDOS is 0.05 eV.
\label{Fig3}}
\end{figure*}


Figure 3(a) shows the PDOS of the CNT on the BN sheet 
when an external electric field is applied in the normal vector direction to the hBN plane.
In all the cases, there is a large difference (more than 2 eV) between the CBM of the BN sheet and $E_F$.
We can conclude that the NFE state does not affect electronic conduction,
and that the conduction occurs only through the CNT in the CNT-based nanodevice. 
In the presence of an external electric field, there is no practical change in the energy levels near 
$E_F$, although small changes are observed in the peak position and height of the PDOS.

To study the doping effect of electrons or holes by a single dopant, we considered 
an atomic impurity (K or Cl) placed between the CNT and the BN sheet, as shown in Fig. 1(d). 
In the calculations, we increased the length of our unit cell ($3 \times 2.50$ \AA) along 
the tube axis. Figures 3(b) and 3(c) show the PDOSs of the K- and Cl-doped CNTs 
on the BN sheet, respectively, under the applied electric field. 
Because of the doping effect, the Fermi levels are upshifted by ${\approx}$0.65 eV (downshifted by ${\approx}$0.43 eV) for the K (Cl) impurity atom. 
Although the electronic energy bands of the BN sheet are modified 
by the electric field, the energy level of NFE state is not very close to $E_F$ under the field 
strength considered. In the case of the K-doped CNT, the CBM of hBN is far from $E_F$ by more than 1 eV above $E_F$ for a field of $-0.2$ eV/\AA.
On the other hand, in the case of the Cl-doped CNT, a field of $-0.2$ eV/\AA~gives rise to the upshift of the valence band maximum (VBM) close to $E_F$.  
According to our Bader charge analysis, the K atom donates 0.85 $e$ to the CNT and the Cl accepts 0.32 $e$ from the CNT.
This result implies that the CNT has mainly ionic interaction with the K atom but it has orbital hybridization as well as ionic interaction with the Cl atom.
We observed that a CNT state hybridized with the Cl impurity state occurs at $E_F$ whereas 
the hBN sheet has no clear coupling (hybridization) with the Cl impurity at $E_F$.
Therefore, the Cl impurity atom may function as a scattering source for the quantum charge transport in the CNT.  
However, we need not worry about the occurrence of unwanted electrical conduction (leakage current) due to the Cl adatom through the BN sheet in the electronic device
because there is no coupling between the Cl adatom and hBN.

We also studied the effect of an external electric field on the electronic properties 
of a Ni-doped CNT on the BN sheet. 
Figure 3(d) shows the PDOS of the Ni-doped CNT on the BN sheet under an external
electric field. In the PDOS, although all the states are slightly modified 
by the electric field, any state of the CBM (NFE state) or the VBM 
in the BN sheet does not touch $E_F$. However, strongly localized PDOS peaks 
originating from the Ni 3$d$ orbital appear near $E_F$. The presence of these peaks show that there is strong 
coupling between the Ni adatom, the CNT and the hBN sheet, which is in sharp 
contrast to the Cl and K doping cases. As mentioned above, the Cl and K dopants do not have impurity-derived BN states near $E_F$.
In the case of Ni-doping, the Bader charge analysis shows that 0.33 $e$ is transferred from the Ni adatom to the CNT.

\begin{figure}[b]
\includegraphics[width=1.0\columnwidth]{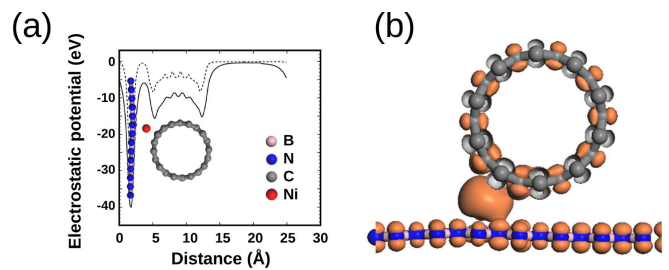}
\caption{(Color online) The local potential averaged over a plane parallel to hBN as a function
of the distance normal to the slab for (a) the undoped and the Ni-doped (10, 0) CNTs.
The dashed line is for the bare CNT on the hBN sheet while the solid line indicates the potential for the Ni-doped CNT.
(b) The local charge density in the energy window where the strong peaks of the Ni PDOS occur, as shown in Figure 3(d).
\label{Fig4}}
\end{figure}

In a single-gate field effect transistor (FET), an electric field is generated 
between the gate and the channel when a gate bias is applied. The BN states 
that are relatively close to $E_F$ due to doping and the electric field maintain the
energy spacing between the VBM (or the  CBM) in the BN sheet and the Fermi level, as in the case of zero 
electric field. The dopant atom (K or Cl) and/or the electric field do not influence
the approach of an electronic band of the BN sheet towards $E_F$. 
Thus, conduction occurs only in the doped CNTs. 
In the light of the above-mentioned results, the BN sheet has emerged 
as a potentially suitable substrate material. However, for the dual-gate FET, 
the Fermi level can shift and equal the CBM of the BN sheet. It is noted that 
the dual-gate FET can control not only the doping level 
(the electronic chemical potential) but also the electric field. For an electric field of $-$0.2 V/\AA~
for the K-doped case, the NFE state in the CBM in the BN sheet is 
close to $E_F$. For an electric field of $+$0.2 V/\AA~electric field for the Cl-doped case, 
the states of the VBM in the BN sheet are close to $E_F$ (0.57 eV below $E_F$).
If the Fermi level equals the CBM of the BN sheet in the dual-gate FET, electronic conduction (leakage current) may occur through the BN sheet.

Finally, we examined the local potentials for the K, Cl, and Ni-doped CNTs on the hBN. 
Figure 4(a) displays the local potential as a function of the distance normal 
to the slab for the undoped CNT on the BN sheet and the field direction 
for the undoped (dashed line) and Ni-doped (solid line) CNTs on the BN sheet 
in the presence of an external electric field. The local potentials between 
the BN sheet and the CNT are the same as in the vacuum space in the right region
in Figure 4(a); this results indicates that electronic screening occurs between the hBN 
sheet and the CNT. The presence of the BN sheet does not alter the fundamental response  
of the undoped CNT. 
We found that, for K- and Cl-doped CNTs, the local potentials between the BN sheet
and the CNT are almost the same as in the vacuum region far from the undoped CNT.
In contrast, for the Ni-doped CNT, the local potential in the region between the hBN sheet and the CNT
is less than those for the other cases by 5.5 eV on the right side of the vacuum region.
In Figure 4(b), the local charge density in the energy window where the strong peaks
of the Ni PDOS occur, as shown in Figure 3(d), exhibits a charge overlap between the Ni-doped CNT and hBN.
This result implies that the nickel atom may play a crucial role as a scattering center 
when the Fermi level is shifted into this energy window via doping or the application of an external electric field. 
In addition, unwanted electrical conduction could occur through the BN sheet 
because there is coupling between the CNT and hBN via the Ni impurity atom.


In summary, we studied the effects of an external electric field and doping 
on the electronic properties of CNTs on the BN sheet, and obtained electronic 
structures (band structure, PDOS and local potential) for three types of atomic impurities
(K, Cl, and Ni) in the presence of the external electric field. We found that 
although the electronic energy bands of the BN sheet in the PDOS are modified 
in accordance with the field strength, its electronic state in the valence or conduction band does not equal
$E_F$ under the field strength considered. Our results suggest that 
the BN sheet can be considered as a potentially suitable substrate material for doped 
CNT-based single-gate FETs. 
However, BN substrates with some defects might exhibit poor performance 
since the imperfections impair electrical conductivity due to residual scattering 
when strong top-gate voltage is applied in applications like dual-gate field effect transistors.


This work was supported by the Priority Research Center Program (2011-0018395) and 
the Basic Science Research Program through MEST/NRF (2010-0007805).



\begin{thebibliography}{26}%
\makeatletter
\providecommand \@ifxundefined [1]{%
 \@ifx{#1\undefined}
}%
\providecommand \@ifnum [1]{%
 \ifnum #1\expandafter \@firstoftwo
 \else \expandafter \@secondoftwo
 \fi
}%
\providecommand \@ifx [1]{%
 \ifx #1\expandafter \@firstoftwo
 \else \expandafter \@secondoftwo
 \fi
}%
\providecommand \natexlab [1]{#1}%
\providecommand \enquote  [1]{``#1''}%
\providecommand \bibnamefont  [1]{#1}%
\providecommand \bibfnamefont [1]{#1}%
\providecommand \citenamefont [1]{#1}%
\providecommand \href@noop [0]{\@secondoftwo}%
\providecommand \href [0]{\begingroup \@sanitize@url \@href}%
\providecommand \@href[1]{\@@startlink{#1}\@@href}%
\providecommand \@@href[1]{\endgroup#1\@@endlink}%
\providecommand \@sanitize@url [0]{\catcode `\\12\catcode `\$12\catcode
  `\&12\catcode `\#12\catcode `\^12\catcode `\_12\catcode `\%12\relax}%
\providecommand \@@startlink[1]{}%
\providecommand \@@endlink[0]{}%
\providecommand \url  [0]{\begingroup\@sanitize@url \@url }%
\providecommand \@url [1]{\endgroup\@href {#1}{\urlprefix }}%
\providecommand \urlprefix  [0]{URL }%
\providecommand \Eprint [0]{\href }%
\providecommand \doibase [0]{http://dx.doi.org/}%
\providecommand \selectlanguage [0]{\@gobble}%
\providecommand \bibinfo  [0]{\@secondoftwo}%
\providecommand \bibfield  [0]{\@secondoftwo}%
\providecommand \translation [1]{[#1]}%
\providecommand \BibitemOpen [0]{}%
\providecommand \bibitemStop [0]{}%
\providecommand \bibitemNoStop [0]{.\EOS\space}%
\providecommand \EOS [0]{\spacefactor3000\relax}%
\providecommand \BibitemShut  [1]{\csname bibitem#1\endcsname}%
\let\auto@bib@innerbib\@empty
\bibitem [{\citenamefont {Marchini}\ \emph {et~al.}(2007)\citenamefont
  {Marchini}, \citenamefont {G\"unther},\ and\ \citenamefont
  {Wintterlin}}]{GRARU}%
  \BibitemOpen
  \bibfield  {author} {\bibinfo {author} {\bibfnamefont {S.}~\bibnamefont
  {Marchini}}, \bibinfo {author} {\bibfnamefont {S.}~\bibnamefont {G\"unther}},
  \ and\ \bibinfo {author} {\bibfnamefont {J.}~\bibnamefont {Wintterlin}},\
  }\href {\doibase 10.1103/PhysRevB.76.075429} {\bibfield  {journal} {\bibinfo
  {journal} {Phys. Rev. B}\ }\textbf {\bibinfo {volume} {76}},\ \bibinfo
  {pages} {075429} (\bibinfo {year} {2007})}\BibitemShut {NoStop}%
\bibitem [{\citenamefont {N'Diaye}\ \emph {et~al.}(2006)\citenamefont
  {N'Diaye}, \citenamefont {Bleikamp}, \citenamefont {Feibelman},\ and\
  \citenamefont {Michely}}]{GRAIR}%
  \BibitemOpen
  \bibfield  {author} {\bibinfo {author} {\bibfnamefont {A.~T.}\ \bibnamefont
  {N'Diaye}}, \bibinfo {author} {\bibfnamefont {S.}~\bibnamefont {Bleikamp}},
  \bibinfo {author} {\bibfnamefont {P.~J.}\ \bibnamefont {Feibelman}}, \ and\
  \bibinfo {author} {\bibfnamefont {T.}~\bibnamefont {Michely}},\ }\href
  {\doibase 10.1103/PhysRevLett.97.215501} {\bibfield  {journal} {\bibinfo
  {journal} {Phys. Rev. Lett.}\ }\textbf {\bibinfo {volume} {97}},\ \bibinfo
  {pages} {215501} (\bibinfo {year} {2006})}\BibitemShut {NoStop}%
\bibitem [{\citenamefont {Lui}\ \emph {et~al.}(2009)\citenamefont {Lui},
  \citenamefont {Liu}, \citenamefont {Mak}, \citenamefont {Flynn},\ and\
  \citenamefont {Heinz}}]{GRAMICA}%
  \BibitemOpen
  \bibfield  {author} {\bibinfo {author} {\bibfnamefont {C.~H.}\ \bibnamefont
  {Lui}}, \bibinfo {author} {\bibfnamefont {L.}~\bibnamefont {Liu}}, \bibinfo
  {author} {\bibfnamefont {K.~F.}\ \bibnamefont {Mak}}, \bibinfo {author}
  {\bibfnamefont {G.~W.}\ \bibnamefont {Flynn}}, \ and\ \bibinfo {author}
  {\bibfnamefont {T.~F.}\ \bibnamefont {Heinz}},\ }\href {\doibase
  10.1038/nature08569} {\bibfield  {journal} {\bibinfo  {journal} {Nature}\
  }\textbf {\bibinfo {volume} {462}},\ \bibinfo {pages} {339} (\bibinfo {year}
  {2009})}\BibitemShut {NoStop}%
\bibitem [{\citenamefont {Berger}\ \emph {et~al.}(2006)\citenamefont {Berger},
  \citenamefont {Song}, \citenamefont {Li}, \citenamefont {Wu}, \citenamefont
  {Brown}, \citenamefont {Naud}, \citenamefont {Mayou}, \citenamefont {Li},
  \citenamefont {Hass}, \citenamefont {Marchenkov}, \citenamefont {Conrad},
  \citenamefont {First},\ and\ \citenamefont {de~Heer}}]{SIC1}%
  \BibitemOpen
  \bibfield  {author} {\bibinfo {author} {\bibfnamefont {C.}~\bibnamefont
  {Berger}}, \bibinfo {author} {\bibfnamefont {Z.}~\bibnamefont {Song}},
  \bibinfo {author} {\bibfnamefont {X.}~\bibnamefont {Li}}, \bibinfo {author}
  {\bibfnamefont {X.}~\bibnamefont {Wu}}, \bibinfo {author} {\bibfnamefont
  {N.}~\bibnamefont {Brown}}, \bibinfo {author} {\bibfnamefont
  {C.}~\bibnamefont {Naud}}, \bibinfo {author} {\bibfnamefont {D.}~\bibnamefont
  {Mayou}}, \bibinfo {author} {\bibfnamefont {T.}~\bibnamefont {Li}}, \bibinfo
  {author} {\bibfnamefont {J.}~\bibnamefont {Hass}}, \bibinfo {author}
  {\bibfnamefont {A.~N.}\ \bibnamefont {Marchenkov}}, \bibinfo {author}
  {\bibfnamefont {E.~H.}\ \bibnamefont {Conrad}}, \bibinfo {author}
  {\bibfnamefont {P.~N.}\ \bibnamefont {First}}, \ and\ \bibinfo {author}
  {\bibfnamefont {W.~A.}\ \bibnamefont {de~Heer}},\ }\href {\doibase
  10.1126/science.1125925} {\bibfield  {journal} {\bibinfo  {journal}
  {Science}\ }\textbf {\bibinfo {volume} {312}},\ \bibinfo {pages} {1191}
  (\bibinfo {year} {2006})}\BibitemShut {NoStop}%
\bibitem [{\citenamefont {Brar}\ \emph {et~al.}(2007)\citenamefont {Brar},
  \citenamefont {Zhang}, \citenamefont {Yayon}, \citenamefont {Ohta},
  \citenamefont {McChesney}, \citenamefont {Bostwick}, \citenamefont
  {Rotenberg}, \citenamefont {Horn},\ and\ \citenamefont {Crommie}}]{SIC2}%
  \BibitemOpen
  \bibfield  {author} {\bibinfo {author} {\bibfnamefont {V.~W.}\ \bibnamefont
  {Brar}}, \bibinfo {author} {\bibfnamefont {Y.}~\bibnamefont {Zhang}},
  \bibinfo {author} {\bibfnamefont {Y.}~\bibnamefont {Yayon}}, \bibinfo
  {author} {\bibfnamefont {T.}~\bibnamefont {Ohta}}, \bibinfo {author}
  {\bibfnamefont {J.~L.}\ \bibnamefont {McChesney}}, \bibinfo {author}
  {\bibfnamefont {A.}~\bibnamefont {Bostwick}}, \bibinfo {author}
  {\bibfnamefont {E.}~\bibnamefont {Rotenberg}}, \bibinfo {author}
  {\bibfnamefont {K.}~\bibnamefont {Horn}}, \ and\ \bibinfo {author}
  {\bibfnamefont {M.~F.}\ \bibnamefont {Crommie}},\ }\href {\doibase
  10.1063/1.2771084} {\bibfield  {journal} {\bibinfo  {journal} {Appl. Phys.
  Lett.}\ }\textbf {\bibinfo {volume} {91}} (\bibinfo {year} {2007}),\
  10.1063/1.2771084}\BibitemShut {NoStop}%
\bibitem [{\citenamefont {Novoselov}\ \emph {et~al.}(2004)\citenamefont
  {Novoselov}, \citenamefont {Geim}, \citenamefont {Morozov}, \citenamefont
  {Jiang}, \citenamefont {Zhang}, \citenamefont {Dubonos}, \citenamefont
  {Grigorieva},\ and\ \citenamefont {Firsov}}]{SIO21}%
  \BibitemOpen
  \bibfield  {author} {\bibinfo {author} {\bibfnamefont {K.}~\bibnamefont
  {Novoselov}}, \bibinfo {author} {\bibfnamefont {A.}~\bibnamefont {Geim}},
  \bibinfo {author} {\bibfnamefont {S.}~\bibnamefont {Morozov}}, \bibinfo
  {author} {\bibfnamefont {D.}~\bibnamefont {Jiang}}, \bibinfo {author}
  {\bibfnamefont {Y.}~\bibnamefont {Zhang}}, \bibinfo {author} {\bibfnamefont
  {S.}~\bibnamefont {Dubonos}}, \bibinfo {author} {\bibfnamefont
  {I.}~\bibnamefont {Grigorieva}}, \ and\ \bibinfo {author} {\bibfnamefont
  {A.}~\bibnamefont {Firsov}},\ }\href {\doibase 10.1126/science.1102896}
  {\bibfield  {journal} {\bibinfo  {journal} {Science}\ }\textbf {\bibinfo
  {volume} {306}},\ \bibinfo {pages} {666} (\bibinfo {year}
  {2004})}\BibitemShut {NoStop}%
\bibitem [{\citenamefont {Zhang}\ \emph {et~al.}(2005)\citenamefont {Zhang},
  \citenamefont {Tan}, \citenamefont {Stormer},\ and\ \citenamefont
  {Kim}}]{SIO22}%
  \BibitemOpen
  \bibfield  {author} {\bibinfo {author} {\bibfnamefont {Y.}~\bibnamefont
  {Zhang}}, \bibinfo {author} {\bibfnamefont {Y.}~\bibnamefont {Tan}}, \bibinfo
  {author} {\bibfnamefont {H.}~\bibnamefont {Stormer}}, \ and\ \bibinfo
  {author} {\bibfnamefont {P.}~\bibnamefont {Kim}},\ }\href {\doibase
  10.1038/nature04235} {\bibfield  {journal} {\bibinfo  {journal} {Nature}\
  }\textbf {\bibinfo {volume} {438}},\ \bibinfo {pages} {201} (\bibinfo {year}
  {2005})}\BibitemShut {NoStop}%
\bibitem [{\citenamefont {Zhang}\ \emph {et~al.}(2008)\citenamefont {Zhang},
  \citenamefont {Brar}, \citenamefont {Wang}, \citenamefont {Girit},
  \citenamefont {Yayon}, \citenamefont {Panlasigui}, \citenamefont {Zettl},\
  and\ \citenamefont {Crommie}}]{SIO23}%
  \BibitemOpen
  \bibfield  {author} {\bibinfo {author} {\bibfnamefont {Y.}~\bibnamefont
  {Zhang}}, \bibinfo {author} {\bibfnamefont {V.~W.}\ \bibnamefont {Brar}},
  \bibinfo {author} {\bibfnamefont {F.}~\bibnamefont {Wang}}, \bibinfo {author}
  {\bibfnamefont {C.}~\bibnamefont {Girit}}, \bibinfo {author} {\bibfnamefont
  {Y.}~\bibnamefont {Yayon}}, \bibinfo {author} {\bibfnamefont
  {M.}~\bibnamefont {Panlasigui}}, \bibinfo {author} {\bibfnamefont
  {A.}~\bibnamefont {Zettl}}, \ and\ \bibinfo {author} {\bibfnamefont {M.~F.}\
  \bibnamefont {Crommie}},\ }\href {\doibase 10.1038/nphys1022} {\bibfield
  {journal} {\bibinfo  {journal} {Nat. Phys.}\ }\textbf {\bibinfo {volume}
  {4}},\ \bibinfo {pages} {627} (\bibinfo {year} {2008})}\BibitemShut {NoStop}%
\bibitem [{\citenamefont {Deshpande}\ \emph {et~al.}(2009)\citenamefont
  {Deshpande}, \citenamefont {Bao}, \citenamefont {Miao}, \citenamefont {Lau},\
  and\ \citenamefont {LeRoy}}]{SIO24}%
  \BibitemOpen
  \bibfield  {author} {\bibinfo {author} {\bibfnamefont {A.}~\bibnamefont
  {Deshpande}}, \bibinfo {author} {\bibfnamefont {W.}~\bibnamefont {Bao}},
  \bibinfo {author} {\bibfnamefont {F.}~\bibnamefont {Miao}}, \bibinfo {author}
  {\bibfnamefont {C.~N.}\ \bibnamefont {Lau}}, \ and\ \bibinfo {author}
  {\bibfnamefont {B.~J.}\ \bibnamefont {LeRoy}},\ }\href {\doibase
  10.1103/PhysRevB.79.205411} {\bibfield  {journal} {\bibinfo  {journal} {Phys.
  Rev. B}\ }\textbf {\bibinfo {volume} {79}},\ \bibinfo {pages} {205411}
  (\bibinfo {year} {2009})}\BibitemShut {NoStop}%
\bibitem [{\citenamefont {Ishigami}\ \emph {et~al.}(2007)\citenamefont
  {Ishigami}, \citenamefont {Chen}, \citenamefont {Cullen}, \citenamefont
  {Fuhrer},\ and\ \citenamefont {Williams}}]{SIO25}%
  \BibitemOpen
  \bibfield  {author} {\bibinfo {author} {\bibfnamefont {M.}~\bibnamefont
  {Ishigami}}, \bibinfo {author} {\bibfnamefont {J.~H.}\ \bibnamefont {Chen}},
  \bibinfo {author} {\bibfnamefont {W.~G.}\ \bibnamefont {Cullen}}, \bibinfo
  {author} {\bibfnamefont {M.~S.}\ \bibnamefont {Fuhrer}}, \ and\ \bibinfo
  {author} {\bibfnamefont {E.~D.}\ \bibnamefont {Williams}},\ }\href {\doibase
  10.1021/nl070613a} {\bibfield  {journal} {\bibinfo  {journal} {Nano Lett.}\
  }\textbf {\bibinfo {volume} {7}},\ \bibinfo {pages} {1643} (\bibinfo {year}
  {2007})}\BibitemShut {NoStop}%
\bibitem [{\citenamefont {Martin}\ \emph {et~al.}(2008)\citenamefont {Martin},
  \citenamefont {Akerman}, \citenamefont {Ulbricht}, \citenamefont {Lohmann},
  \citenamefont {Smet}, \citenamefont {Von~Klitzing},\ and\ \citenamefont
  {Yacoby}}]{REDM}%
  \BibitemOpen
  \bibfield  {author} {\bibinfo {author} {\bibfnamefont {J.}~\bibnamefont
  {Martin}}, \bibinfo {author} {\bibfnamefont {N.}~\bibnamefont {Akerman}},
  \bibinfo {author} {\bibfnamefont {G.}~\bibnamefont {Ulbricht}}, \bibinfo
  {author} {\bibfnamefont {T.}~\bibnamefont {Lohmann}}, \bibinfo {author}
  {\bibfnamefont {J.~H.}\ \bibnamefont {Smet}}, \bibinfo {author}
  {\bibfnamefont {K.}~\bibnamefont {Von~Klitzing}}, \ and\ \bibinfo {author}
  {\bibfnamefont {A.}~\bibnamefont {Yacoby}},\ }\href {\doibase
  10.1038/nphys781} {\bibfield  {journal} {\bibinfo  {journal} {Nat. Phys}\
  }\textbf {\bibinfo {volume} {4}},\ \bibinfo {pages} {144} (\bibinfo {year}
  {2008})}\BibitemShut {NoStop}%
\bibitem [{\citenamefont {Zhang}\ \emph {et~al.}(2009)\citenamefont {Zhang},
  \citenamefont {Brar}, \citenamefont {Girit}, \citenamefont {Zettl},\ and\
  \citenamefont {Crommie}}]{REDM1}%
  \BibitemOpen
  \bibfield  {author} {\bibinfo {author} {\bibfnamefont {Y.}~\bibnamefont
  {Zhang}}, \bibinfo {author} {\bibfnamefont {V.~W.}\ \bibnamefont {Brar}},
  \bibinfo {author} {\bibfnamefont {C.}~\bibnamefont {Girit}}, \bibinfo
  {author} {\bibfnamefont {A.}~\bibnamefont {Zettl}}, \ and\ \bibinfo {author}
  {\bibfnamefont {M.~F.}\ \bibnamefont {Crommie}},\ }\href {\doibase
  10.1038/NPHYS1365} {\bibfield  {journal} {\bibinfo  {journal} {Nat. Phys.}\
  }\textbf {\bibinfo {volume} {5}},\ \bibinfo {pages} {722} (\bibinfo {year}
  {2009})}\BibitemShut {NoStop}%
\bibitem [{\citenamefont {Gannett}\ \emph {et~al.}(2011)\citenamefont
  {Gannett}, \citenamefont {Regan}, \citenamefont {Watanabe}, \citenamefont
  {Taniguchi}, \citenamefont {Crommie},\ and\ \citenamefont {Zettl}}]{IMMO}%
  \BibitemOpen
  \bibfield  {author} {\bibinfo {author} {\bibfnamefont {W.}~\bibnamefont
  {Gannett}}, \bibinfo {author} {\bibfnamefont {W.}~\bibnamefont {Regan}},
  \bibinfo {author} {\bibfnamefont {K.}~\bibnamefont {Watanabe}}, \bibinfo
  {author} {\bibfnamefont {T.}~\bibnamefont {Taniguchi}}, \bibinfo {author}
  {\bibfnamefont {M.~F.}\ \bibnamefont {Crommie}}, \ and\ \bibinfo {author}
  {\bibfnamefont {A.}~\bibnamefont {Zettl}},\ }\href {\doibase
  10.1063/1.3599708} {\bibfield  {journal} {\bibinfo  {journal} {Appl. Phys.
  Lett.}\ }\textbf {\bibinfo {volume} {98}} (\bibinfo {year} {2011}),\
  10.1063/1.3599708}\BibitemShut {NoStop}%
\bibitem [{\citenamefont {Dean}\ \emph {et~al.}()\citenamefont {Dean},
  \citenamefont {Young}, \citenamefont {Meric}, \citenamefont {Lee},
  \citenamefont {Wang}, \citenamefont {Sorgenfrei}, \citenamefont {Watanabe},
  \citenamefont {Taniguchi}, \citenamefont {Kim}, \citenamefont {Shepard},\
  and\ \citenamefont {Hone}}]{HQGRA}%
  \BibitemOpen
  \bibfield  {author} {\bibinfo {author} {\bibfnamefont {C.~R.}\ \bibnamefont
  {Dean}}, \bibinfo {author} {\bibfnamefont {A.~F.}\ \bibnamefont {Young}},
  \bibinfo {author} {\bibfnamefont {I.}~\bibnamefont {Meric}}, \bibinfo
  {author} {\bibfnamefont {C.}~\bibnamefont {Lee}}, \bibinfo {author}
  {\bibfnamefont {L.}~\bibnamefont {Wang}}, \bibinfo {author} {\bibfnamefont
  {S.}~\bibnamefont {Sorgenfrei}}, \bibinfo {author} {\bibfnamefont
  {K.}~\bibnamefont {Watanabe}}, \bibinfo {author} {\bibfnamefont
  {T.}~\bibnamefont {Taniguchi}}, \bibinfo {author} {\bibfnamefont
  {P.}~\bibnamefont {Kim}}, \bibinfo {author} {\bibfnamefont {K.~L.}\
  \bibnamefont {Shepard}}, \ and\ \bibinfo {author} {\bibfnamefont
  {J.}~\bibnamefont {Hone}},\ }\href@noop {} {\ }\BibitemShut {NoStop}%
\bibitem [{\citenamefont {Xue}\ \emph {et~al.}(2011)\citenamefont {Xue},
  \citenamefont {Sanchez-Yamagishi}, \citenamefont {Bulmash}, \citenamefont
  {Jacquod}, \citenamefont {Deshpande}, \citenamefont {Watanabe}, \citenamefont
  {Taniguchi}, \citenamefont {Jarillo-Herrero},\ and\ \citenamefont
  {Leroy}}]{STM1}%
  \BibitemOpen
  \bibfield  {author} {\bibinfo {author} {\bibfnamefont {J.}~\bibnamefont
  {Xue}}, \bibinfo {author} {\bibfnamefont {J.}~\bibnamefont
  {Sanchez-Yamagishi}}, \bibinfo {author} {\bibfnamefont {D.}~\bibnamefont
  {Bulmash}}, \bibinfo {author} {\bibfnamefont {P.}~\bibnamefont {Jacquod}},
  \bibinfo {author} {\bibfnamefont {A.}~\bibnamefont {Deshpande}}, \bibinfo
  {author} {\bibfnamefont {K.}~\bibnamefont {Watanabe}}, \bibinfo {author}
  {\bibfnamefont {T.}~\bibnamefont {Taniguchi}}, \bibinfo {author}
  {\bibfnamefont {P.}~\bibnamefont {Jarillo-Herrero}}, \ and\ \bibinfo {author}
  {\bibfnamefont {B.~J.}\ \bibnamefont {Leroy}},\ }\href@noop {} {\bibfield
  {journal} {\bibinfo  {journal} {Nat. Mater.}\ }\textbf {\bibinfo {volume}
  {10}},\ \bibinfo {pages} {282} (\bibinfo {year} {2011})}\BibitemShut
  {NoStop}%
\bibitem [{\citenamefont {Decker}\ \emph {et~al.}(2011)\citenamefont {Decker},
  \citenamefont {Wang}, \citenamefont {Brar}, \citenamefont {Regan},
  \citenamefont {Tsai}, \citenamefont {Wu}, \citenamefont {Gannett},
  \citenamefont {Zettl},\ and\ \citenamefont {Crommie}}]{STM2}%
  \BibitemOpen
  \bibfield  {author} {\bibinfo {author} {\bibfnamefont {R.}~\bibnamefont
  {Decker}}, \bibinfo {author} {\bibfnamefont {Y.}~\bibnamefont {Wang}},
  \bibinfo {author} {\bibfnamefont {V.~W.}\ \bibnamefont {Brar}}, \bibinfo
  {author} {\bibfnamefont {W.}~\bibnamefont {Regan}}, \bibinfo {author}
  {\bibfnamefont {H.-Z.}\ \bibnamefont {Tsai}}, \bibinfo {author}
  {\bibfnamefont {Q.}~\bibnamefont {Wu}}, \bibinfo {author} {\bibfnamefont
  {W.}~\bibnamefont {Gannett}}, \bibinfo {author} {\bibfnamefont
  {A.}~\bibnamefont {Zettl}}, \ and\ \bibinfo {author} {\bibfnamefont {M.~F.}\
  \bibnamefont {Crommie}},\ }\href {\doibase 10.1021/nl2005115} {\bibfield
  {journal} {\bibinfo  {journal} {Nano Lett.}\ }\textbf {\bibinfo {volume}
  {11}},\ \bibinfo {pages} {2291} (\bibinfo {year} {2011})}\BibitemShut
  {NoStop}%
\bibitem [{\citenamefont {Watanabe}\ \emph {et~al.}(2004)\citenamefont
  {Watanabe}, \citenamefont {Taniguchi},\ and\ \citenamefont {Kanda}}]{BG}%
  \BibitemOpen
  \bibfield  {author} {\bibinfo {author} {\bibfnamefont {K.}~\bibnamefont
  {Watanabe}}, \bibinfo {author} {\bibfnamefont {T.}~\bibnamefont {Taniguchi}},
  \ and\ \bibinfo {author} {\bibfnamefont {H.}~\bibnamefont {Kanda}},\ }\href
  {\doibase 10.1038/nmat1134} {\bibfield  {journal} {\bibinfo  {journal} {Nat.
  Mater.}\ }\textbf {\bibinfo {volume} {3}},\ \bibinfo {pages} {404} (\bibinfo
  {year} {2004})}\BibitemShut {NoStop}%
\bibitem [{\citenamefont {Blase}\ \emph {et~al.}(1995)\citenamefont {Blase},
  \citenamefont {Rubio}, \citenamefont {Louie},\ and\ \citenamefont
  {Cohen}}]{NFE}%
  \BibitemOpen
  \bibfield  {author} {\bibinfo {author} {\bibfnamefont {X.}~\bibnamefont
  {Blase}}, \bibinfo {author} {\bibfnamefont {A.}~\bibnamefont {Rubio}},
  \bibinfo {author} {\bibfnamefont {S.~G.}\ \bibnamefont {Louie}}, \ and\
  \bibinfo {author} {\bibfnamefont {M.~L.}\ \bibnamefont {Cohen}},\ }\href
  {\doibase 10.1103/PhysRevB.51.6868} {\bibfield  {journal} {\bibinfo
  {journal} {Phys. Rev. B}\ }\textbf {\bibinfo {volume} {51}},\ \bibinfo
  {pages} {6868} (\bibinfo {year} {1995})}\BibitemShut {NoStop}%
\bibitem [{\citenamefont {Kresse}\ and\ \citenamefont
  {Furthm\"uller}(1996)}]{VASP}%
  \BibitemOpen
  \bibfield  {author} {\bibinfo {author} {\bibfnamefont {G.}~\bibnamefont
  {Kresse}}\ and\ \bibinfo {author} {\bibfnamefont {J.}~\bibnamefont
  {Furthm\"uller}},\ }\href {\doibase 10.1103/PhysRevB.54.11169} {\bibfield
  {journal} {\bibinfo  {journal} {Phys. Rev. B}\ }\textbf {\bibinfo {volume}
  {54}},\ \bibinfo {pages} {11169} (\bibinfo {year} {1996})}\BibitemShut
  {NoStop}%
\bibitem [{\citenamefont {Kresse}\ and\ \citenamefont {Joubert}(1999)}]{PAW}%
  \BibitemOpen
  \bibfield  {author} {\bibinfo {author} {\bibfnamefont {G.}~\bibnamefont
  {Kresse}}\ and\ \bibinfo {author} {\bibfnamefont {D.}~\bibnamefont
  {Joubert}},\ }\href {\doibase 10.1103/PhysRevB.59.1758} {\bibfield  {journal}
  {\bibinfo  {journal} {Phys. Rev. B}\ }\textbf {\bibinfo {volume} {59}},\
  \bibinfo {pages} {1758} (\bibinfo {year} {1999})}\BibitemShut {NoStop}%
\bibitem [{\citenamefont {Ceperley}\ and\ \citenamefont {Alder}(1980)}]{LDACA}%
  \BibitemOpen
  \bibfield  {author} {\bibinfo {author} {\bibfnamefont {D.~M.}\ \bibnamefont
  {Ceperley}}\ and\ \bibinfo {author} {\bibfnamefont {B.~J.}\ \bibnamefont
  {Alder}},\ }\href {\doibase 10.1103/PhysRevLett.45.566} {\bibfield  {journal}
  {\bibinfo  {journal} {Phys. Rev. Lett.}\ }\textbf {\bibinfo {volume} {45}},\
  \bibinfo {pages} {566} (\bibinfo {year} {1980})}\BibitemShut {NoStop}%
\bibitem [{\citenamefont {Jo}\ \emph {et~al.}(2002)\citenamefont {Jo},
  \citenamefont {Kim},\ and\ \citenamefont {Lee}}]{KCNT1}%
  \BibitemOpen
  \bibfield  {author} {\bibinfo {author} {\bibfnamefont {C.}~\bibnamefont
  {Jo}}, \bibinfo {author} {\bibfnamefont {C.}~\bibnamefont {Kim}}, \ and\
  \bibinfo {author} {\bibfnamefont {Y.~H.}\ \bibnamefont {Lee}},\ }\href
  {\doibase 10.1103/PhysRevB.65.035420} {\bibfield  {journal} {\bibinfo
  {journal} {Phys. Rev. B}\ }\textbf {\bibinfo {volume} {65}},\ \bibinfo
  {pages} {035420} (\bibinfo {year} {2002})}\BibitemShut {NoStop}%
\bibitem [{\citenamefont {Hansson}\ and\ \citenamefont
  {Stafstr\"om}(2005)}]{KCNT2}%
  \BibitemOpen
  \bibfield  {author} {\bibinfo {author} {\bibfnamefont {A.}~\bibnamefont
  {Hansson}}\ and\ \bibinfo {author} {\bibfnamefont {S.}~\bibnamefont
  {Stafstr\"om}},\ }\href {\doibase 10.1103/PhysRevB.72.125420} {\bibfield
  {journal} {\bibinfo  {journal} {Phys. Rev. B}\ }\textbf {\bibinfo {volume}
  {72}},\ \bibinfo {pages} {125420} (\bibinfo {year} {2005})}\BibitemShut
  {NoStop}%
\bibitem [{\citenamefont {Pan}\ \emph {et~al.}(2006)\citenamefont {Pan},
  \citenamefont {Feng},\ and\ \citenamefont {Lin}}]{ClCNT1}%
  \BibitemOpen
  \bibfield  {author} {\bibinfo {author} {\bibfnamefont {H.}~\bibnamefont
  {Pan}}, \bibinfo {author} {\bibfnamefont {Y.~P.}\ \bibnamefont {Feng}}, \
  and\ \bibinfo {author} {\bibfnamefont {J.~Y.}\ \bibnamefont {Lin}},\ }\href
  {http://stacks.iop.org/0953-8984/18/i=22/a=017} {\bibfield  {journal}
  {\bibinfo  {journal} {J. Phys.: Condens. Matter}\ }\textbf {\bibinfo {volume}
  {18}},\ \bibinfo {pages} {5175} (\bibinfo {year} {2006})}\BibitemShut
  {NoStop}%
\bibitem [{\citenamefont {Medeiros}\ \emph {et~al.}(2010)\citenamefont
  {Medeiros}, \citenamefont {Mascarenhas}, \citenamefont {de~Brito~Mota},\ and\
  \citenamefont {de~Castilho}}]{ClCNT2}%
  \BibitemOpen
  \bibfield  {author} {\bibinfo {author} {\bibfnamefont {P.~V.~C.}\
  \bibnamefont {Medeiros}}, \bibinfo {author} {\bibfnamefont {A.~J.~S.}\
  \bibnamefont {Mascarenhas}}, \bibinfo {author} {\bibfnamefont
  {F.}~\bibnamefont {de~Brito~Mota}}, \ and\ \bibinfo {author} {\bibfnamefont
  {C.~M.~C.}\ \bibnamefont {de~Castilho}},\ }\href
  {http://stacks.iop.org/0957-4484/21/i=48/a=485701} {\bibfield  {journal}
  {\bibinfo  {journal} {Nanotechnology}\ }\textbf {\bibinfo {volume} {21}},\
  \bibinfo {pages} {485701} (\bibinfo {year} {2010})}\BibitemShut {NoStop}%
\bibitem [{\citenamefont {Liu}\ \emph {et~al.}(2007)\citenamefont {Liu},
  \citenamefont {Gurel}, \citenamefont {Morris}, \citenamefont {Murray},
  \citenamefont {Zhitkovich}, \citenamefont {Kane},\ and\ \citenamefont
  {Hurt}}]{NiCNT}%
  \BibitemOpen
  \bibfield  {author} {\bibinfo {author} {\bibfnamefont {X.}~\bibnamefont
  {Liu}}, \bibinfo {author} {\bibfnamefont {V.}~\bibnamefont {Gurel}}, \bibinfo
  {author} {\bibfnamefont {D.}~\bibnamefont {Morris}}, \bibinfo {author}
  {\bibfnamefont {D.}~\bibnamefont {Murray}}, \bibinfo {author} {\bibfnamefont
  {A.}~\bibnamefont {Zhitkovich}}, \bibinfo {author} {\bibfnamefont
  {A.}~\bibnamefont {Kane}}, \ and\ \bibinfo {author} {\bibfnamefont
  {R.}~\bibnamefont {Hurt}},\ }\href {\doibase 10.1002/adma.200602696}
  {\bibfield  {journal} {\bibinfo  {journal} {Adv. Mater.}\ }\textbf {\bibinfo
  {volume} {19}},\ \bibinfo {pages} {2790} (\bibinfo {year}
  {2007})}\BibitemShut {NoStop}%
\end{thebibliography}

%

\end{document}